\documentclass[english,aps,prl,preprint]{revtex4-1}

\usepackage[T1]{fontenc}
\usepackage[latin9]{inputenc}
\usepackage{color}
\usepackage{units}
\usepackage{amsmath}
\usepackage{amssymb}
\usepackage{wasysym}
\usepackage{graphicx}
\usepackage{esint}

\makeatletter
 
 \@ifundefined{textcolor}{}
 {%
   \definecolor{BLACK}{gray}{0}
   \definecolor{WHITE}{gray}{1}
   \definecolor{RED}{rgb}{1,0,0}
   \definecolor{GREEN}{rgb}{0,1,0}
   \definecolor{BLUE}{rgb}{0,0,1}
   \definecolor{CYAN}{cmyk}{1,0,0,0}
   \definecolor{MAGENTA}{cmyk}{0,1,0,0}
   \definecolor{YELLOW}{cmyk}{0,0,1,0}
 }

\@ifundefined{definecolor}
 {\usepackage{color}}{}

\makeatother

\usepackage{babel}
\begin{document}
\color{black}

\title{Spatial cooperativity in microchannel flows of soft jammed materials:
A mesoscopic approach}

\author{Alexandre Nicolas, Jean-Louis Barrat}

\affiliation{Laboratoire Interdisciplinaire de Physique, Université Joseph Fourier
Grenoble, CNRS UMR 5588, BP 87, 38402 Saint-Martin d'Hères, France}

\date{December,10$^{th}$ 2012}
\begin{abstract}
The flow of amorphous solids results from a combination of elastic
deformation and local structural rearrangements, which induce non-local
elastic deformations. These elements are incorporated into a mechanically-consistent
mesoscopic model of interacting elastoplastic blocks. We investigate
the specific case of channel flow with numerical simulations, paying
particular attention to situations of strong confinement. We find
that the simple picture of plastic events embedded in an elastic matrix
successfully accounts for manifestations of spatial cooperativity.
Shear rate fluctuations are observed in seemingly quiescent regions,
and the velocity profiles in confined flows at high applied pressure
deviate from those expected in the absence of nonlocal effects,
in agreement with experimental data. However, we suggest a different
physical origin for the large deviations observed when walls
have rough surfaces, associated with  {}``bumps'' of the particles
against the asperities of the walls. 
\end{abstract}

\pacs{47.57.Qk, 83.80.Ab, 62.20.fq}

\maketitle
Shear waves are damped in liquids, and propagate in elastic solids.
This well-established distinction has major implications in the field
of seismology  \citep{Shearer1999}, but also has bearing on the peculiar nonlocal
rheology of soft jammed/glassy materials, which share solid (elastic)
and liquid (flow) properties.

In fact, the flow of these materials bears notable similarities with
earthquakes: it features a solid-like behavior at rest and local yielding
above a given applied stress. Yielding is characterized by the emergence
of local {}``shear transformations'' \citep{Argon1979} involving
a few particles \citep{Schall2007}, associated with a local fluidization
of the material. These structural rearrangements, hereafter named
plastic events, induce long-range deformations. The microscopic details
vary to some extent with the particular nature of the material. In
the case of foams, they are identified as T1 events \citep{PRINCEN1985}
in which the local change of  neighbors is mediated by an unstable
stage with four bubbles sharing one vertex. The robustness of the
scenario for an extremely wide range of materials is striking.
Ample evidence of the local plastic events and their long-range effects
is indeed provided both by experiments using diverse materials \citep{Schall2007,Amon2012,Nichol2012}
and by simulations \citep{Falk1998,Lemaitre2007}.

Turning to the specific case of channel flow, the existence of long-range
interactions render any purely local approach questionable in a system
featuring an inherently inhomogeneous shear and specific boundary
conditions due to the walls. Convincing experimental evidence disproves
the very \emph{existence} of\emph{ }a local constitutive relation
$\sigma=\sigma\left(\dot{\gamma}\right)$ \citep{Goyon2008a}, where
$\dot{\gamma}$ and $\sigma$ are the local shear rate and shear stress.
The presence of nonlocal effects  highlights the effect of
spatial cooperativity. These observations were rationalized in \citep{Bocquet2009,Goyon2008a}
by introducing a diffusion equation for the local fluidity $f=\nicefrac{\dot{\gamma}}{\sigma}$:
\begin{equation}
\xi^{2}\Delta f-\left(f-f_{bulk}(\sigma)\right)=0,\label{eq:Fluidity_diff}
\end{equation}
 Here $\xi$ is a cooperativity length, and $f_{bulk}$ is the local
bulk fluidity, measured in a homogeneous, simple shear situation.
Eq.\ref{eq:Fluidity_diff} has been tested with considerable success
\citep{Jop2012,Goyon2010}. However, the approach ignores the fluctuating
nature of plastic deformation, and relies on experimental measurements
of the boundary conditions.

Mesoscopic models \citep{Baret2002,Picard2005,Martens2012} offer a computationally-efficient way to recover
part of the complexity of the dynamics, while leaving behind microscopic
details. Most importantly, they allow us to test our understanding
of the physical processes involved in the flow, and identify relevant
parameters for its description \citep{Martens2012}. However, direct
comparison with experimental data is scarce.

In this Letter, we build upon previous work \citep{Picard2005,Martens2012}
to develop a simple, but mechanically consistent, 2D tensorial mesoscopic
model that incorporates the phenomenology reported above, as well
as (coarsened) convection. For the first time such a model is used
to simulate a channel flow, and its predictions are compared to 
experimental data.

The channel is modeled as a rectangular elastic matrix, which is spatially
discretized into a regular lattice of square-shaped blocks. Under
the assumption of isotropy and incompressibility, Hooke's law states
that the deviatoric stress tensor $\boldsymbol{\sigma}$ is proportional
to the local strain tensor $\boldsymbol{\epsilon}$, 
\begin{equation}
\sigma_{xx}=\mu\left(\text{\ensuremath{\epsilon}}_{xx}-\text{\ensuremath{\epsilon}}_{yy}\right)=2\mu\epsilon_{xx}\ ;\ \ \sigma_{xy}=2\mu\text{\ensuremath{\epsilon}}_{xy}.\label{Hooke}
\end{equation}
 $\mu$ is the shear modulus, and \emph{x} and \emph{y} are the flow
and gradient directions, respectively. The system covers the domain
$\left(x,y\right)\in[0,L_{x}]\times[0,L_{y}]$, with $L_{x}$ and
$L_{y}$ the channel length and width. The initial response of the
material confined in a channel to an applied pressure gradient $\nabla p\, e_{x}$
is therefore given by: $\sigma_{xy}\left(x,y,t=0\right)=\nabla p\ \left(y-\nicefrac{L_{y}}{2}\right)$
and $\sigma_{xx}\left(x,y,t=0\right)=0$.

As soon as a yielding criterion is met in a block, the block undergoes
a plastic event, during which the stored elastic energy is dissipated.
Within the framework of eigenstrain theory \citep{Mura1987}, the
elastic strain $\boldsymbol{\epsilon}$ is converted into a plastic
eigenstrain $\boldsymbol{\epsilon^{pl}}$. Physically, this corresponds
to the change of reference elastic configuration following particle
rearrangement. The conversion
takes a finite time, due to dissipative effects.

Since the block is embedded in an elastic matrix, the localized eigenstrain
$\boldsymbol{\epsilon^{pl}}$ induces an elastic field $\boldsymbol{\sigma}^{(1)}\left(r\right)=\int\boldsymbol{\mathcal{G}}\left(r,r^{\prime}\right)\cdot2\mu\boldsymbol{\epsilon^{pl}}\left(r^{\prime}\right)dr^{\prime}$,
where the elastic propagator $\boldsymbol{\mathcal{G}}$ depends on
the boundary conditions. The assumption of an underlying elastic behavior
is bolstered by the similarity of the induced field with experimental
and numerical observations \citep{Schall2007,Lemaitre2007}, all of
which exhibit a quadrupolar angular dependence and a $\Vert r-r^{\prime}\Vert^{-d}$-like
decay in \emph{d} dimensions. Combining the different elements, one gets the equation of
evolution of a block: 
\begin{equation}
\partial_{t}\boldsymbol{\sigma}\left(r\right)=\boldsymbol{\dot{\Sigma}^{ext}}\left(r\right)+\int\boldsymbol{\mathcal{G}}\left(r,r^{\prime}\right)\cdot2\mu\boldsymbol{\dot{\epsilon}^{pl}}\left(r^{\prime}\right)d^{2}r^{\prime}\label{eq: DynEq}
\end{equation}
 Here, $\boldsymbol{\dot{\epsilon}^{pl}}=\frac{1}{\tau}\boldsymbol{\epsilon}$
if the block is plastic, $\boldsymbol{\dot{\epsilon}^{pl}}=0$ otherwise,
and $\tau$ is the characteristic timescale for the release of the
stored elastic stress. $\boldsymbol{\dot{\Sigma}^{ext}}$ is the time-derivative
of the stress response of a purely elastic material to the same applied
conditions; therefore $\boldsymbol{\dot{\Sigma}^{ext}}=0$ if the
flow is pressure-driven. Numerically, at each time step, Eq.\ref{eq: DynEq}
is solved in Fourier space, on a finer grid where, for accuracy, each
block is divided into four subcells.

There is obviously some arbitrariness regarding the choice of the
yielding criterion and the duration of plastic events. The corresponding
switching rates are denoted by $l\left(\sigma\right)$ and $e\left(\sigma\right)$,
\[
\text{elastic regime}\overset{l\left(\sigma\right)}{\underset{e\left(\sigma\right)}{\rightleftharpoons}}\text{plastic event}.
\]
 The probability to yield $l\left(\sigma\right)$ is set to $l\left(\sigma\right)=\Theta\left(\sigma-\sigma_{\mu y}\right)\exp\left(\frac{\sigma-\sigma_{y}}{x_{l}}\right)\tau^{-1}$,
where $\Theta$ is the Heaviside function and $\sigma=\Vert\boldsymbol{\sigma}\Vert$,
so as to mimic an activated process of \emph{non-cooperative} origin,
associated with a material-dependent intensive parameter $x_{l}$.
Imposing a finite critical stress $\sigma_{\mu y}$, below which no
plastic event can occur, is crucial for the existence of a macroscopic
yield stress $\sigma\left(\dot{\gamma}\rightarrow0\right)\neq0$.
Also note that the von Mises yield criterion is recovered in the
limit $x_{l}\rightarrow0$. The above expression  accounts
for the long-time relaxation of a pre-sheared material, observed even
in the case of (athermal) granular materials \citep{Hartley2003}.
We assume that elasticity is restored when the plastic deformation
rate becomes low, and  choose: $e\left(\sigma\right)=\exp\left(\frac{\sigma_{\mu y}-\sigma}{x_{e}}\right)\tau^{-1}$,
where we have introduced another material parameter, $x_{e}$.

Walls are modeled by imposing no-slip boundary conditions at $y=0$
and $y=L_{y}$. This results in a lengthy expression for the propagator, which we defer to a longer
publication, along with details of the derivation. It is however noteworthy
that, for a given eigenstrain $\boldsymbol{\dot{\epsilon}^{pl}}$,
the elastic stress that is released locally is up to 35\% larger than
in the bulk if the plastic event occurs near a wall.

Finally, a coarsened picture of convection is implemented, whereby lines of blocks in the flow direction (instead of individual blocks)
are incrementally shifted.

\begin{figure}
\includegraphics[width=190pt]{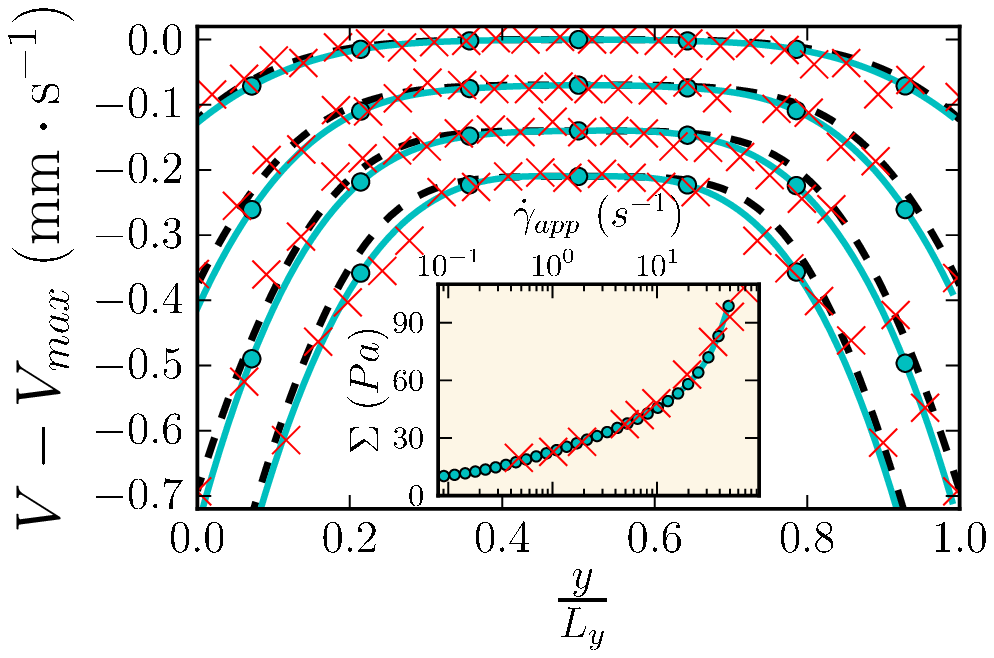}

\caption{\label{fig:Vel_Goyon_smooth_walls}Velocity profiles in the microchannel.
The curves are shifted for clarity. (${\color{red}\times}$)
Experimental measurements by Goyon \emph{et al.} \citep{Goyon2010}
for (from top to bottom) $\sigma_{w}=45,\,60,\,75,\,91\,\unit{Pa}$,
i.e., $\sigma_{w}=0.75,\:1.0,\,1.25,\,1.52$ in model units, in a
$112\,\unit{\mu m}$-wide smooth channel; (${\color{cyan}\bullet}$)
numerical results for $N_{y}=7$. (The solid line is a guide to the
eye); \emph{dashed black line}: predicted velocity profile on the basis of
the macroscopic flow curve $\Sigma\left(\dot{\gamma}_{app}\right)$,
shown in the \emph{inset} graph ((${\color{red}\times}$) experimental data from \citep{Goyon2010};
 (${\color{cyan}\bullet}$) numerical results in a large system, $N_{y}=64$).}
\end{figure}

We choose units of time and stress such that $\tau=1,\,\mu=1$ and
we set $\sigma_{y}=1$. The remaining model parameters, $\sigma_{\mu y}$,
$x_{l}$, and $x_{e}$, are now fitted by comparing the predictions
of the model in simple shear flow, i.e., $\sigma\left(t=0\right)=0$
and $\dot{\Sigma}^{ext}=\mu\dot{\gamma}_{app}$ in Eq.\ref{eq: DynEq},
where $\dot{\gamma}_{app}$ is the applied shear rate, with the macroscopic
rheology measurements collected by Goyon and co-workers \citep{Goyon2010,Goyon2008a}
for an oil-in-water emulsion of average radius $6.5\,\unit{\mu m}$
, which follows a Herschel-Bulkley law $\sigma=\sigma_{0}+A\dot{\gamma}^{n},\, n\approx0.5$.
The inset of Fig.\ref{fig:Vel_Goyon_smooth_walls} shows that the
experimental data can satisfactorily be reproduced with our model
by using the following parameters: $\sigma_{\mu y}$=0.17, $x_{l}=0.249$,
and $x_{e}=1.66$. Note that the model units of stress and time have
been appropriately rescaled.

Cooperativity is a general feature of the flow of amorphous solids,
regardless of the flow geometry and the driving. Channel flow, however,
is specific in that 

(i) the non-locality of the stress redistribution couples streamlines
subject to \emph{different} shear stresses.

(ii) the presence of a wall, whether it be rough or smooth, may create
a specific surface rheology, different from that in the bulk. 

In practice, these effects are of primary importance for confined
flows in microchannels, but they are not intrinsically caused by confinement.

Let us first consider Point (i). To estimate its importance, we introduce
a dimensionless number, the Babel number $\mathcal{B}a\equiv\delta f/f_{bulk}(\sigma)$,
where $\delta f=f-f_{bulk}$ is the deviation from the expected fluidity
profile owing to cooperative effects between regions subject to different
driving forces. Within the theoretical framework of fluidity diffusion,
Eq. \ref{eq:Fluidity_diff}, and for a fluid obeying a Herschel Bulkley
relation in the bulk, it can be shown that  $\mathcal{B}a=\left(\xi\frac{\Vert\nabla\sigma\Vert}{\sigma-\sigma_{0}}\right)^{2}=\left(\xi\frac{\Vert\nabla p\Vert}{\sigma-\sigma_{0}}\right)^{2}$
(in regions where the locally imposed stress $\sigma$ is larger than
the yield stress $\sigma_{0}$). Large deviations from bulk behavior
are expected when the stress is close to the yield value. A striking
consequence of this first point was very recently unveiled by Jop \emph{et al.}
\citep{Jop2012} in a system similar to that of Goyon
\emph{et al.} \citep{Goyon2010}. They analyzed experimental data
for  an oil-in-water emulsion in an almost two-dimensional
microchannel flow,
and showed that the seemingly quiescent, rigid plug is in fact subject
to finite shear rate fluctuations $\delta\dot{\gamma}\left(x,y\right)=\sqrt{\left\langle \dot{\gamma}\left(x,y\right)^{2}\right\rangle -\left\langle \dot{\gamma}\left(x,y\right)\right\rangle ^{2}}$,
where the brackets denote time averages. This clearly points to a
nonlocal effect of the sheared regions.

\begin{figure}
\includegraphics[width=190pt]{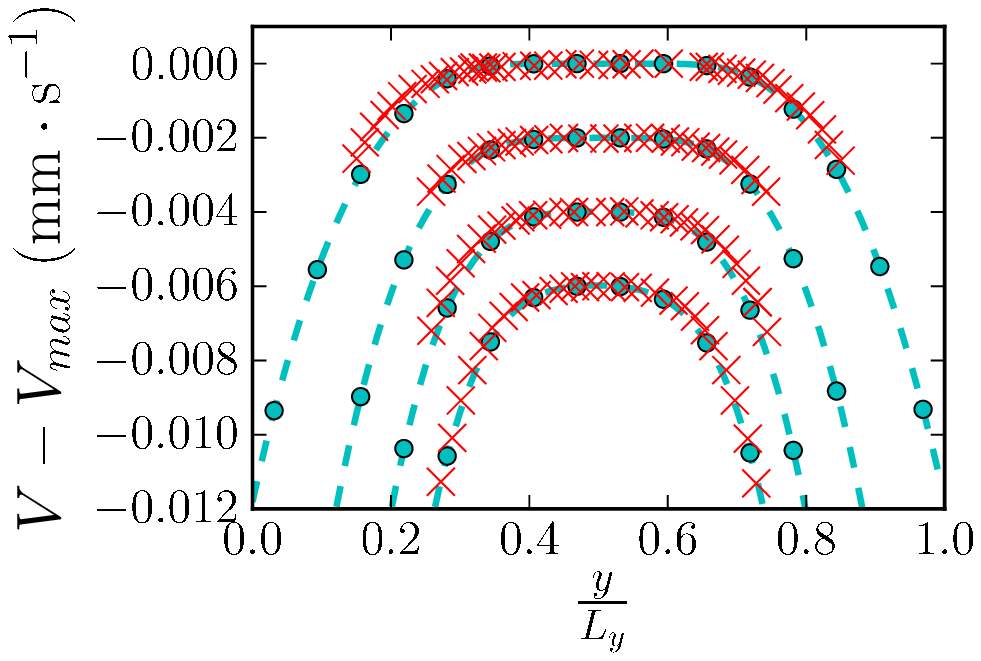}\caption{\label{fig:Jop_vel_profile}Offset velocity as a function of the channel
crosswise coordinate, for stresses at the wall $\sigma_{w}=$$141\,\unit{Pa}$,
$188\,\unit{Pa}$, $235\,\unit{Pa}$, $282\,\unit{Pa}$, corresponding
to $\sigma_{w}=0.36,\,0.48,\,0.60,\,0.72$ in model units, from top
to bottom. $\left({\color{red}\times}\right)$ Experimental data collected
by Jop\emph{ et al.} \citep{Jop2012} (channel width: $225\,\mathrm{\mu m}$),
$\left({\color{cyan}\bullet}\right)$ numerical results for a channel
of $N_{y}=16$ blocks crosswise. The curves are offset with respect
to each other for clarity.}
\end{figure}

\begin{figure}
\includegraphics[width=190pt]{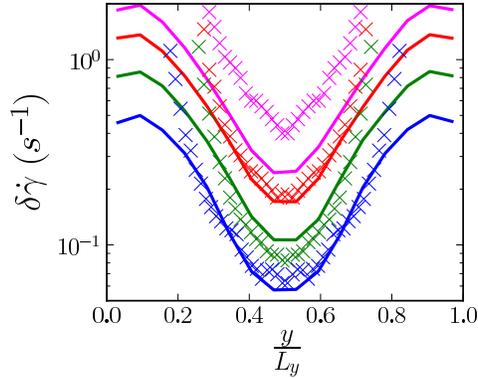}

\caption{\label{fig:Jop_Fluct_08}Shear rate fluctuations $\delta\dot{\gamma}\left(y\right)$
(averaged along the \emph{x}-direction), for $\sigma_{w}=$$141\,\unit{Pa}$,
$188\,\unit{Pa}$, $235\,\unit{Pa}$, $282\,\unit{Pa}$ (identical
to Fig.\ref{fig:Jop_vel_profile}), from bottom to top. $\left({\color{red}\times}\right)$
Experimental data collected by Jop \emph{et al.} \citep{Jop2012},
(\emph{solid lines}) numerical results for $N_{y}=16$.}
\end{figure}

We now compare the model results\footnote{Videos are provided as Supplemental Material.} directly to experimental data:
Fig.\ref{fig:Jop_vel_profile} shows the velocity profiles
\footnote{We have discarded the curves corresponding to the lowest two applied
pressures in \citep{Jop2012}, because we are not sure of the accuracy
of the measurements of the lowest shear rate fluctuations.%
}, and Fig.\ref{fig:Jop_Fluct_08} shows
the corresponding shear rate fluctuation profiles, for a numerical channel of transverse size
$N_{y}=16$ blocks.

 Although the parameters of the model have been constrained
only weakly by adjusting the macroscopic flow curve, semi-quantitative
agreement is observed in regions far from the walls - apart from the
large discrepancy at the highest applied pressure. The discrepancies
in the vicinity of the walls will be considered below. It is interesting
to note that the fitted channel size provides an estimate for the
linear size $N_{\diameter}$ of an elastoplastic block measured in
particle diameters, $N_{\diameter}\approx2$.

Not only shear rate fluctuations, but also the average shear rate,
elude a local, or bulk, description, when $\mathcal{B}a$
gets large. This paradigm-shifting fact is the major result of experimental
observations by Goyon and co-workers in microchannel flows of emulsions
with either smooth or rough walls \citep{Goyon2008a}.

In the case of smooth walls, deviations from the macroscopic bulk
constitutive relations are comparatively weak. In contrast to 
strain rate fluctuations, they become perceptible only at high applied
pressures in confined geometries,  when the stress at the
wall is larger than some channel-size dependent critical value $\sigma_{w}^{\star}$.
Under these conditions, the overall fluidity of the system is enhanced.
Consistently with our expression of the Babel number, the most prominent
discrepancies with the bulk predictions are observed near the edges
of the plug, which become more rounded. Accordingly, cooperativity
can soften the material so that it flows even below the macroscopic
yield stress. This aspect is corroborated by numerical simulations
\citep{Chaudhuri2012}, and is expected within the framework of the
fluidity diffusion equation \citep{Bocquet2009}. Fig.\ref{fig:Vel_Goyon_smooth_walls}
shows that our model also satisfactorily captures these features.
For the critical wall stresses $\sigma_{w}^{\star}$ to coincide between
the model and the experiments, the transverse size of the numerical
channel must be set between 6 and 10 blocks, which again corresponds
to $N_{\diameter}\approx2$. Besides, for such transverse microchannel
sizes, the maximal velocity displays large oscillations in time, as
has often been reported experimentally\citep{Isa2009}.

\begin{figure}
\includegraphics[width=190pt]{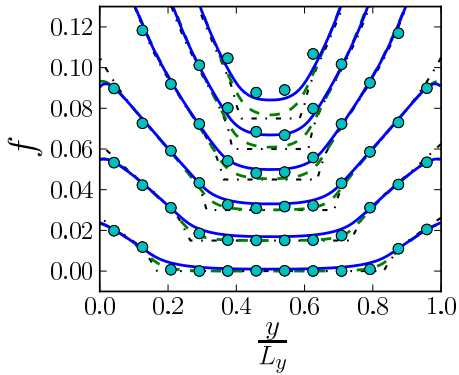}

\caption{\label{fig:Fluidity_diffusion}Fluidity profiles for $N_{y}=12$.
Filled circles: numerical results, dashed green line: solution of
Eq.\ref{eq:Fluidity_diff} with $\xi\left(\dot{\gamma}\right)=0.03702$,
solid blue line: solution of Eq.\ref{eq:Fluidity_diff} with $\xi\left(\dot{\gamma}\right)=0.01146\,\dot{\gamma}^{-0.25}$.
The thin dash-dotted lines represent the bulk fluidity $f_{bulk}$.}
\end{figure}

Let us now turn to
a comparison with a description in terms of fluidity diffusion, Eq.
\ref{eq:Fluidity_diff}. To solve Eq. \ref{eq:Fluidity_diff}, the
shear-rate dependence of the  length $\xi$ must be specified,
and two boundary conditions (BC) are required. Regarding the BC,
we impose $f\left(y=0\right)=f\left(y=L_{y}\right)$, and set the
fluidity at a point close to the wall to the value measured in simulations.
For the fluidity dependence of $\xi$, two cases are studied in Fig.\ref{fig:Fluidity_diffusion}:
either no dependence, i.e., $\xi=\xi_{0}$, following Goyon \emph{et
al. }\citep{Goyon2008a}\emph{, }or a power-law dependence, $\xi\left(\dot{\gamma}\right)=\xi_{0}\dot{\gamma}^{-0.25}$,
where $\dot{\gamma}$ is the product of the local shear stress and
fluidity, as derived in Ref. \citep{Bocquet2009} in the limit $\dot{\gamma}\rightarrow0$,
and in reasonable agreement with the data of Ref. \citep{Jop2012}.
In both cases, $\xi_{0}$ is adjusted by a least square minimization.
Overall, the power-law dependence provides a closer fit of our numerical
results, especially at higher applied pressures. However, neither
assumption concerning $\xi\left(\dot{\gamma}\right)$ was able to
provide a perfect fit, a defect that we ascribe to the approximation
of long-range interactions by a diffusive term, and to the neglect
of fluidity fluctuations.

Now, when rough walls are substituted for smooth walls, the situation
differs widely. Much larger deviations are  observed, even at
lower Babel numbers than above. The  discrepancies caused by
the change of the surface roughness of the walls point to the prevalence
the specific surface rheology, point (ii) above. Further evidence
comes from the fact that deviations are particularly large close to
the walls. The no-slip BC at the wall imposed
in our model are insufficient to account for these large deviations.
One is therefore led to conclude that either (a) the roughness  of the wall alters the  structure
of the material in its vicinity or (b) slip along the wall generates
a stress field in the system, as particles constantly bump into wall
asperities.

Since the  observed deviations are not restricted to the direct
vicinity of the (rough) walls, we examine the second possibility here,
namely slip along the wall as a source of mechanical noise. Goyon \emph{et al.} \citep{Goyon2008a} measured
wall slip both for smooth and rough surfaces, the latter having asperities
of a characteristic lengthscale of $1\,\unit{\mu m}$, whereas the
emulsion is made of $\sim6\,\mu m$-diameter droplets. In the case
of smooth surfaces, asperities are too small to deform the particles
significantly, and the presence of wall slip is therefore unlikely
to induce considerable change as compared to the no-slip case, which
validates our approach. On the contrary, wall slip along a rough surfaces
brings about {}``bumps'' (or deformations) of the particles into
surface asperities. Similarly to bulk plastic events, these {}``bumps''
are expected to induce an elastic field in the material. This
scenario has the potential to explain why deviations are not always
observed in nominally similar conditions: Seth and co-workers \citep{Seth2012},
for instance, used a different surface preparation protocol to obtain
rough surfaces and did not detect any nonlinearity of the flow in
the vicinity of the rough walls, but, quite interestingly, they also
reported that wall slip  remained negligible in that case.

Our model cannot render the mechanical effects of wall slip along
a rough surface without further input. Nevertheless, fictitious plastic
events can artificially be added along the walls to mimic the impact
of {}``bumps''. Concretely, we now model both walls as lines of
blocks and select a fraction ($\nicefrac{1}{3}$) of these blocks at random as mechanical
noise sources, that is to say, they shall release a constant plastic
strain $\dot{\epsilon}_{xy}^{fict\, pl}\left(\approx 5 \right)$ by unit time. It should
be remarked that the procedure does on no account violate mechanical
equilibrium. The resulting local flow curves, shown as Supplemental Material, are qualitatively similar 
to those obtained experimentally in Ref. \citep{Goyon2008a}.
Let us also remark that shear rate profiles
in the presence of fictitious plastic events do not flatten in the
vicinity of the walls (\emph{data
not shown}), as  in Fig.\ref{fig:Jop_Fluct_08},
and are therefore more compatible with the experimental data.

In conclusion, we have refined and extended a mesoscopic model based
on a default elastic behavior and finite-time plastic events to simulate
a confined channel flow. We have realized the first direct comparison
of such a model with experimental data, paying particular attention
to manifestations of spatial cooperativity. It has been found that,
the model  accounts for the presence of
shear-rate fluctuations in seemingly quiescent regions, and the local
deviations from the macroscopic flow curve when the inhomogeneity
of the driving, quantified by the Babel number, is large. The comparison
with experimental data also provided us with an estimate of the size
of an elastoplastic block, $N_{\diameter}\approx2$, roughly in agreement
with experimentally-measured values of the size of a shear transformation
\citep{Schall2007}. These successes of the mesoscopic model are encouraging
for further studies of statistical aspects of flow in complex fluids.
The description in terms of fluidity diffusion is reasonable, but
imperfect.

The much larger deviations observed in the presence of rough walls,
on the other hand, are not described by a model that imposes a fixed
velocity at the walls. This points to the contribution of another
physical mechanism to the deviations. We hypothesize that the bumps
of the particles against the surface asperities due to wall slip,
and the long-range elastic deformations they induce, may be at its
source. Further experimental and numerical studies of wall rheology
will be needed to quantify this mechanism.

\begin{acknowledgments}
AN \& JLB thank K. Martens, D. Rodney, L. Bocquet and P. Chaudhuri
for interesting discussions, and A. Colin for sending the experimental
data. AN acknowledges a fruitful discussion with R. Besseling, JLB
is supported by Institut Universitaire de France and by grant ERC-2011-ADG20110209.
\end{acknowledgments}

\bibliographystyle{apsrev4-1}
%
 
 \part*{Supplementary material}

\section*{Fictitious plastic events as sources of mechanical noise}

\begin{figure}
\includegraphics[width=190pt]{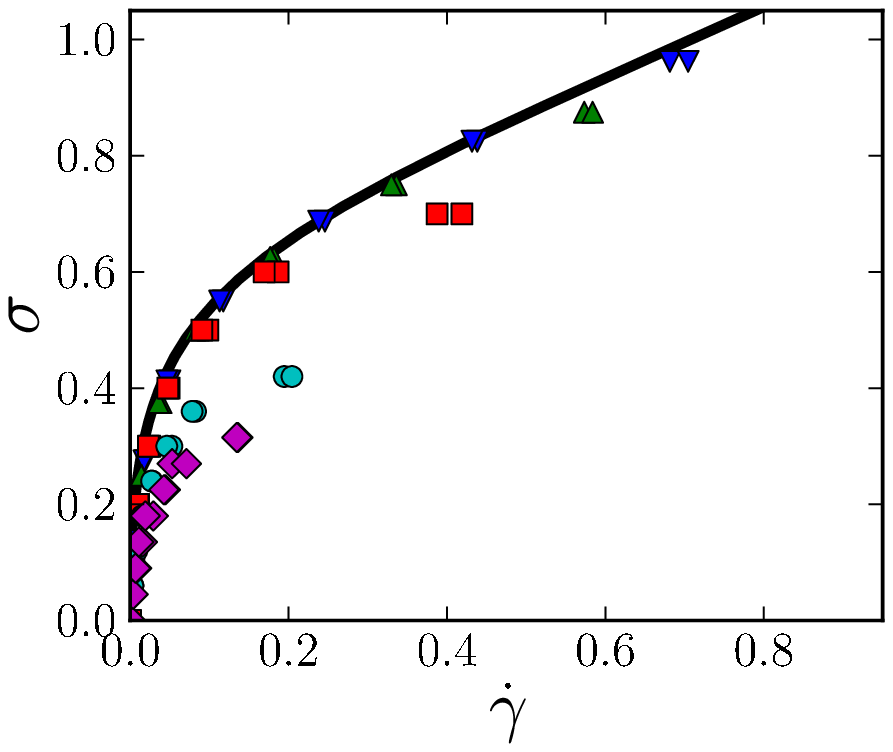}

\caption{\label{fig:Noisy_local_flow_curve}Local shear rate $\sigma\left(y\right)$
\emph{vs }local shear rate $\dot{\gamma}\left(y\right)$ (averaged
on streamlines $y=cst$) in the microchannel, when fictitious mechanical
noise sources of intensity $\dot{\epsilon}_{xy}^{fict\, pl}=\pm4.5$
are added on a fraction ($\nicefrac{1}{3}$) of blocks on the wall
lines. $\sigma_{w}$=$\left({\color{magenta}\blacklozenge}\right)$
$0.36$, $\left({\color{cyan}\bullet}\right)$ $0.48$, $\left({\color{red}\blacksquare}\right)$$0.8$,
$\left({\color{green}\blacktriangle}\right)$1.0, $\left({\color{blue}\blacktriangledown}\right)$$1.1$
in model units. Solid line: macroscopic flow curve. }
\end{figure}

To mimick the impact of the {}``bumps'' into wall asperities induced
by wall slip, we introduce fictitious plastic events along the wall.
More precisely, we now model both walls as lines of blocks and select
a fraction of these blocks at random as mechanical noise sources,
that is to say, they shall release a constant plastic strain $\dot{\epsilon}_{xy}^{fict\, pl}$
by unit time. Fig.\ref{fig:Noisy_local_flow_curve} shows the resulting
local flow curves for a fraction of fixed mechanical noise sources
of $\nicefrac{1}{3}$ picked at random, and $\dot{\epsilon}_{xy}^{fict\, pl}=\pm4.5$.

\section*{Videos showing the spatiotemporal dynamics}

Files \emph{Video0p75.mpg} and \emph{Video0p48.mpg} are videos representing
the succession of plastic events in a channel flow.

The system flows along the horizontal direction, towards the right,
and the flow is confined between two horizontal walls, at the top
and at the bottom. A plastic event is depicted as a coloured square
that gradually fades into black as the plastic event comes to an end. 

The color of the plastic event represents its principal direction,
\emph{ie} the yielding angle: Blocks that yield in the macroscopic
shear direction, ie $\dot{\epsilon}_{xx}^{pl}=0$ and $\dot{\epsilon}_{xy}^{pl}\neq0$
, are painted in red, while plastic events with $\dot{\epsilon}_{xx}^{pl}\neq0$
and $\dot{\epsilon}_{xy}^{pl}=0$ are represented in blue. A color
gradient is applied between these limiting cases.

\emph{Video0p75.mpg} shows a system of $N_{y}=7$ blocks in the crosswise
direction and $N_{x}=64$ blocks in the streamwise direction (only
half of the horizontal axis is shown). The stress imposed at the wall
is $\sigma_{w}=0.75$ in model units, and the other model parameters
are identical to those used in the paper. The video covers a time
span $\Delta t=33.5$ in model units.

\emph{Video0p48.mpg} shows a system of $N_{y}=16$ blocks in the crosswise
direction and $N_{x}=64$ blocks in the streamwise direction (only
half of the horizontal axis is shown). The stress imposed at the wall
is $\sigma_{w}=0.75$ in model units, and the other model parameters
are identical to those used in the paper. The video covers a time
span $\Delta t=28.3$ in model units.

\end{document}